\documentclass[aps,prl,reprint,superscriptaddress,amsmath,amssymb,floatfix,longbibliography]{revtex4-2}
\usepackage[T1]{fontenc}
\usepackage[utf8]{inputenc}
\usepackage{amsmath,amssymb,bm,mathtools}
\usepackage{graphicx}
\usepackage{booktabs,array}
\usepackage{microtype}
\usepackage{hyperref}
\hypersetup{colorlinks=true,citecolor=blue,linkcolor=blue,urlcolor=blue}
\newcommand{\Tr}{\operatorname{Tr}}
\newcommand{\D}{\mathcal{D}}
\newcommand{\cL}{\mathcal{L}}

\newcommand{\opt}{\operatorname{opt}}
\newcommand{\ii}{\mathrm{i}}
\newcommand{\ee}{\mathrm{e}}
\newcommand{\dd}{\mathrm{d}}

\begin{document}
\title{Symmetry-constrained order hierarchy of connected optical responses in a five-level Rydberg–EIT medium}
\author{Lintian Luo}
\affiliation{School of Physics, East China Normal University, Shanghai 200241, China}
\author{Weixuan Gu}
\affiliation{School of Physics, East China Normal University, Shanghai 200241, China}
\author{Yan Li}
\email{yli@phy.ecnu.edu.cn}
\affiliation{School of Physics, East China Normal University, Shanghai 200241, China}
\affiliation{Chongqing Key Laboratory of Precision Optics, Chongqing Institute of East China Normal University, Chongqing 401120, China}
\begin{abstract}
Determining the connected many-body correlations required for a nonlinear optical response of a given order is a central challenge in interacting open quantum systems. We establish a symmetry-based selection rule for connected optical responses in a driven-dissipative Rydberg--EIT model. Under the specified Lindblad conditions, local phase covariance and zero-probe factorization imply that an $n$-atom connected contribution to the root optical coherence is forbidden below probe-amplitude order $2n-1$. Consequently, optical-response coefficients through fifth order involve at most three-atom connected clusters, without assuming weak interatomic interactions. Direct finite-cluster calculations confirm the predicted lowest allowed orders, while an illustrative Maxwell-propagation calculation shows that triplet contributions can partially oppose pair-induced polarization rotation. Our results establish a rigorous connection between optical perturbation order and the connected many-body sectors that can contribute to it.
\end{abstract}
\maketitle

The optical response of an interacting medium generally depends on correlations among its constituent particles. Yet the Hilbert-space dimension of an open many-body system grows exponentially with particle number, making it difficult to know how many-body correlations must be retained to predict a nonlinear response of a prescribed order. Truncations based on mean-field or low-order correlation closures are useful, but do not by themselves establish whether omitted connected sectors are forbidden or merely small. This raises a basic question: \emph{can symmetry determine the highest connected-cluster order needed for a given optical perturbation order?}

Rydberg electromagnetically induced transparency (EIT) offers a concrete setting in which strong van der Waals interactions produce optical nonlinearities \cite{Fleischhauer2005EIT,Saffman2010Rydberg,Lukin2001Blockade,Gorshkov2011PhotonInteractions,Peyronel2012QuantumNonlinear,Firstenberg2013AttractivePhotons,Sevincli2011Nonlocal,Sinclair2019CrossKerr}. With two probe components, the resulting optical coherences also control polarization rotation \cite{Budker2002NMOR,Wang2006EITRotation,Li2006ZeemanRotation,Pustelny2006NMOR}. Unlike an earlier effective Raman-dressing treatment of wave-mixing-induced rotation \cite{Luo2026PRA}, we retain the auxiliary level and both wave-mixing couplings explicitly in a five-level Lindblad model; related wave-mixing and polarization studies provide further context \cite{Zhou2017WaveMixing,Zhu2018PropagationBlockade,Yin2026RamanNMOR}.

We show that local phase covariance, combined with factorization when a non-root probe is switched off, requires an $n$-atom M\"obius-connected root-optical response to start no earlier than probe-amplitude order $2n-1$. Thus fifth-order coefficients contain no connected contribution from four or more atoms even when interactions are retained exactly within each finite cluster. Direct one-, two-, and three-atom calculations test the allowed orders, and an illustrative Maxwell propagation shows how pair and triplet sectors compete in the output polarization rotation.

\begin{figure*}[t]
\centering\includegraphics[width=0.94\textwidth]{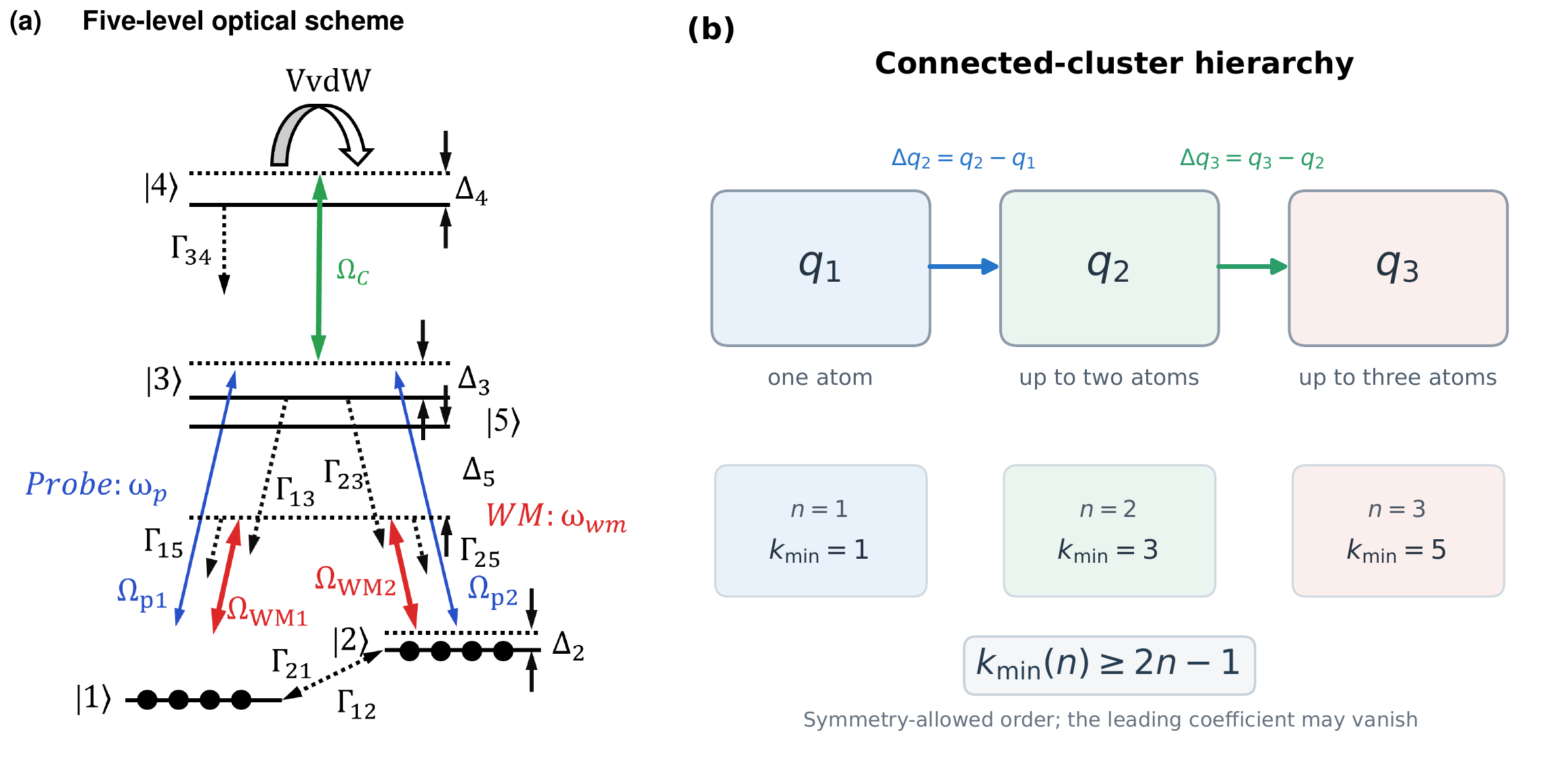}
\caption{Five-level optical scheme and connected-cluster hierarchy. (a) The two circular probe components couple $|1\rangle$ and $|2\rangle$ to $|3\rangle$; the control field couples $|3\rangle$ to the Rydberg state $|4\rangle$, and the two WM components couple the lower states to the auxiliary state $|5\rangle$. The dashed arrows indicate ground-state incoherent transfer; the decay rates are labeled, and $V_{ij}$ denotes the Rydberg interaction. (b) The single-, pair-, and triplet-truncated root responses and their increments. The first symmetry-allowed orders are $1$, $3$, and $5$, respectively.}
\label{fig:model}
\end{figure*}

For atom $i$, let $P_{ai}=|a\rangle_i\langle a|$ and let the internal states $|1\rangle$, $|2\rangle$, $|3\rangle$, $|4\rangle$, and $|5\rangle$ denote two lower Zeeman states, an optical intermediate state, a Rydberg state, and a wave-mixing (WM) auxiliary state, respectively. In the $^{85}\mathrm{Rb}$ realization used here, the states are $|1\rangle=|5S_{1/2},F=3,m_F=1\rangle$, $|2\rangle=|5S_{1/2},F=3,m_F=-1\rangle$, $|3\rangle=|5P_{3/2},F=4,m_F=0\rangle$, $|4\rangle=|68S_{1/2},F=3,m_F=1\rangle$, and $|5\rangle=|5P_{1/2},F=3,m_F=0\rangle$ \cite{Steck2025Rb85,Safronova2004Rb}. For a finite atom set $A$, the Hamiltonian is
\begin{align}
H_A={}&\sum_{i\in A}(H_{i,\mathrm{diag}}+H_{i,\mathrm{WM}}\nonumber\\
&+H_{i,c}+H_{i,p})+\sum_{i<j\in A}V_{ij}P_{4i}P_{4j},\label{eq:H}\\
H_{i,\mathrm{diag}}={}&-\Delta_2P_{2i}-\Delta_3P_{3i}\nonumber\\
&-\Delta_4P_{4i}-\Delta_5P_{5i},\\
H_{i,\mathrm{WM}}={}&\Omega_{\mathrm{WM},1}|5\rangle_i\langle1|\nonumber\\
&+\Omega_{\mathrm{WM},2}|5\rangle_i\langle2|+\mathrm{H.c.},\\
H_{i,c}={}&\Omega_c|4\rangle_i\langle3|+\mathrm{H.c.},\\
H_{i,p}={}&\sum_{\alpha=1}^{2}\bigl(z_{i\alpha}|3\rangle_i\langle\alpha|\nonumber\\
&+\bar z_{i\alpha}|\alpha\rangle_i\langle3|\bigr).
\end{align}

The five-level coupling scheme and connected-cluster structure are illustrated in Fig.~\ref{fig:model}.

In the numerical model, $V_{ij}=-C_6/r_{ij}^{6}$, with the Rydberg interaction parameters and atomic data taken from established calculations and data compilations \cite{Singer2005RydbergC6,Sibalic2017ARC,Steck2025Rb85,Safronova2004Rb}. The density operator obeys the GKSL equation \cite{Lindblad1976,GKS1976}
\begin{align}
\dot\rho_A&=\cL_A(\rho_A)\nonumber\\
&=-\ii[H_A,\rho_A]+\sum_{i\in A,\mu}\D[J_{i\mu}](\rho_A),\\
\D[J](X)&=JXJ^\dagger-\tfrac12\{J^\dagger J,X\}.
\end{align}
The local jumps can be written compactly as
\begin{equation}
\begin{gathered}
J_{3a}=\sqrt{\Gamma_3/2}\,|a\rangle\langle3|,\quad
J_{5a}=\sqrt{\Gamma_5/2}\,|a\rangle\langle5|,\\
J_{34}=\sqrt{\Gamma_4}\,|3\rangle\langle4|,\quad
J_{ab}=\sqrt{\Gamma_{ab}}\,|a\rangle\langle b|,\\
a=1,2,\qquad(a,b)=(1,2),(2,1).
\end{gathered}
\label{eq:jumps}
\end{equation}
An optional low-manifold dephasing channel $J_{\phi}=\sqrt{\pi\gamma_\phi/1000}(P_1-P_2)$ may be included. The order bound below applies to this jump-operator set under the conditions $\Gamma_3,\Gamma_5,\Gamma_{12},\Gamma_{21}>0$, $\Gamma_4\geq0$, $\Omega_c\neq0$, and at least one nonzero WM coupling. No additional zero-probe drive into $|3\rangle$ or $|4\rangle$ is allowed.

For each nonempty subset $B\subseteq A$, let $\rho_B$ be the normalized steady state of the isolated subcluster $B$. The M\"obius-connected operator is
\begin{equation}
g_A=\sum_{\pi\in\Pi(A)}(-1)^{|\pi|-1}(|\pi|-1)!\bigotimes_{B\in\pi}\rho_B,
\label{eq:mobius}
\end{equation}
where $\Pi(A)$ is the set of partitions of $A$ and tensor factors are placed in the physical atom order. This partition-lattice construction is the operator analogue of the cumulant framework \cite{Kubo1962Cumulant,Speed1983PartitionLattices}; its inversion is governed by the M\"obius function of the partition lattice \cite{Rota1964Mobius}. For two and three atoms,
\begin{align}
g_{12}={}&\rho_{12}-\rho_1\otimes\rho_2,\\
g_{123}={}&\rho_{123}-\rho_1\otimes\rho_2\otimes\rho_3\nonumber\\
&-g_{12}\otimes\rho_3-g_{13}\otimes\rho_2\nonumber\\
&-\rho_1\otimes g_{23}.
\end{align}
For the root atom $1$, define
\begin{align}
K_{12}&=\Tr_2g_{12},\quad K_{123}=\Tr_{23}g_{123},\\
\opt(X)&=(X_{31},X_{32})^T.
\end{align}
The pair and triplet contributions to the root's cluster-truncated effective density matrix are
\begin{align}
\mathcal C_2&=4\pi\nu\int_{r_c}^{r_{\max}}r^2K_{12}(r)\,\dd r,\\
\mathcal C_3&=\frac{\nu^2}{2}\int_{\mathcal{D}}\dd^3r_2\,\dd^3r_3\,K_{123}(\mathbf r_2,\mathbf r_3),
\end{align}
where $\nu$ is the atomic density and $\mathcal D$ imposes the short-range cutoff on all three pair separations. We use
\begin{equation}
\rho_{\mathrm{eff},1}=\rho_1,\quad
\rho_{\mathrm{eff},2}=\rho_1+\mathcal C_2,\quad
\rho_{\mathrm{eff},3}=\rho_1+\mathcal C_2+\mathcal C_3,
\end{equation}
and define $q_1=\opt(\rho_{\mathrm{eff},1})$, $q_2=\opt(\rho_{\mathrm{eff},2})$, and $q_3=\opt(\rho_{\mathrm{eff},3})$. The optical pair and triplet increments are denoted by $\Delta q_2=q_2-q_1$ and $\Delta q_3=q_3-q_2$, respectively. These are density-cluster truncations, not exact many-body density matrices.

\begin{figure*}[t]
\centering\includegraphics[width=0.94\textwidth]{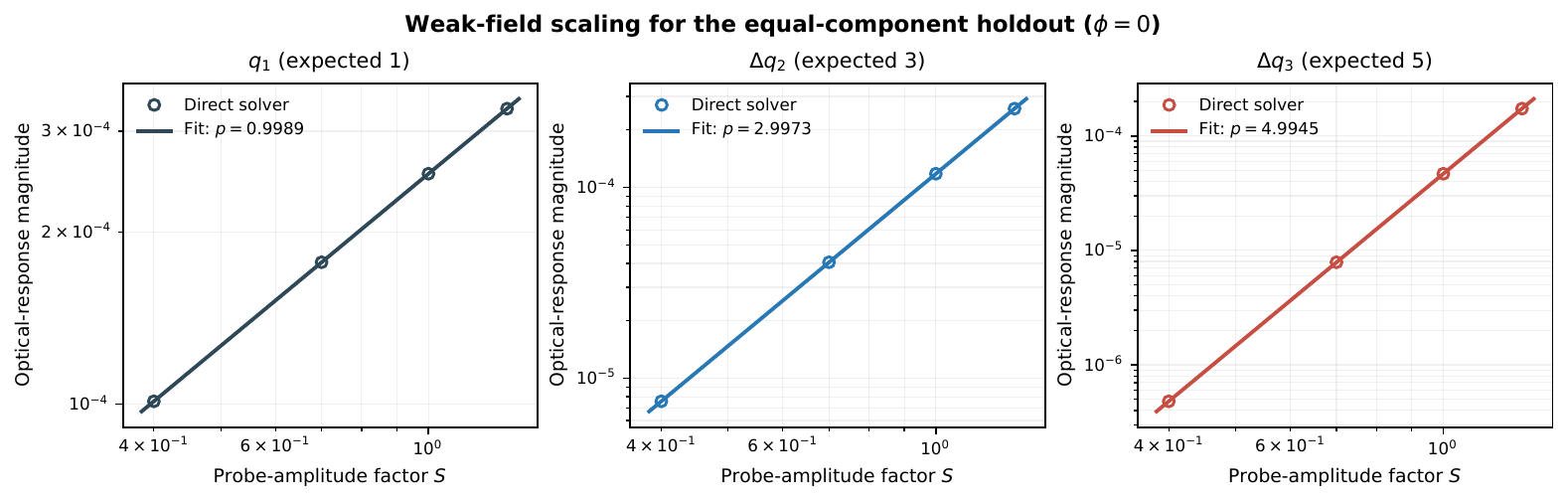}
\caption{Weak-field scaling of the directly calculated single-atom response $q_1$ and the connected pair and triplet increments $\Delta q_2=q_2-q_1$ and $\Delta q_3=q_3-q_2$ for the equal-component validation mode with relative phase $\phi=0$. The fitted finite-window exponents are $0.9989$, $2.9973$, and $4.9945$, respectively, close to the symmetry-allowed orders $1$, $3$, and $5$.}
\label{fig:scaling}
\end{figure*}

The symmetry underlying the order hierarchy is a local internal phase covariance, rather than a spatial symmetry. Define $Q_i=P_{3i}+P_{4i}$ and $U_i(\theta_i)=\exp(\ii\theta_iQ_i)$, so that $|3\rangle$ and $|4\rangle$ carry unit charge while $|1\rangle$, $|2\rangle$, and $|5\rangle$ are neutral. At zero probe, the WM couplings act within the neutral manifold, the control field couples states with the same unit charge, and the diagonal detunings and vdW interaction commute with every $Q_i$. Each specified jump operator changes by at most a phase under $U_i$, leaving its Lindblad dissipator invariant. The probe amplitudes transform as $z_{i\alpha}\mapsto\ee^{\ii\theta_i}z_{i\alpha}$. The finite-cluster Liouvillian therefore has a local $U(1)^{|A|}$ covariance, which imposes an independent charge-selection rule at each atom and constrains the allowed probe-amplitude orders.

For each fixed finite cluster and fixed control and WM fields, detunings, dissipative rates, and cluster geometry, the trace-constrained steady state is locally real analytic in the real and imaginary parts of the probe amplitudes. Its optical response therefore has a locally convergent expansion in homogeneous polynomials of all orders. The Neumann representation in Supplementary Sec.~S3 makes the coefficient structure explicit: every Taylor coefficient is independent of the probe amplitudes and is determined recursively by the zero-probe Liouvillian and the probe-coupling superoperators. For a common two-component probe, the coefficients at degree $2m+1$ can be assembled into a $2\times[(m+2)(m+1)]$ matrix; the first three have dimensions $2\times2$, $2\times6$, and $2\times12$. Local $U(1)$ covariance determines which monomials in these expansions can have nonzero coefficients.

\emph{Selection rule.---} Under the jump-operator and nonzero-coupling assumptions stated above, every nonzero Taylor monomial in the root-optical response of an $n$-atom M\"obius-connected cluster has total probe-amplitude degree
\begin{equation}
\boxed{k\geq 1+2(n-1)=2n-1.}\label{eq:bound}
\end{equation}
The bound is a first symmetry-allowed order; it does not guarantee that the coefficient at that order is nonzero for every parameter set.

The proof has two ingredients. First, the zero-probe stationary state is supported on $\mathcal H_L^{\otimes n}$, where $\mathcal H_L=\mathrm{span}\{|1\rangle,|2\rangle,|5\rangle\}$. On this low manifold, the jumps $J_{51}$, $J_{52}$, $J_{12}$, and $J_{21}$ and their adjoints generate the full matrix algebra. The restricted semigroup is therefore primitive. This gives a unique faithful zero-probe stationary state, an algebraically simple zero eigenvalue, and an invertible zero-probe generator on the trace-zero subspace. Analytic perturbation of this finite-dimensional trace-constrained steady-state equation then gives a locally unique real-analytic steady branch for every fixed finite cluster.

Second, if the probe on a non-root atom $j$ is set to zero, the stationary state factorizes as $\rho_A=\rho_{A\setminus\{j\}}\otimes\sigma_j$, where $\sigma_j$ is the low-manifold stationary state. Since $P_{4j}$ vanishes on this support, every interaction term $V_{ij}P_{4i}P_{4j}$ involving atom $j$ also vanishes there. M\"obius inversion consequently gives $g_A=0$ whenever any non-root probe is switched off. Local covariance requires the root optical coherence to carry net charge $+1$ and every non-root atom to carry zero net charge. Thus a nonzero monomial contains at least one root probe factor and, for each of the $n-1$ non-root atoms, at least one unconjugated and one conjugated probe factor. This yields Eq.~(\ref{eq:bound}). The complete supporting arguments are provided in Supplementary Sections S1--S6. The order bound is established directly for each finite cluster, prior to carrying out the spatial cluster integrals.

\begin{figure*}[t]
\centering\includegraphics[width=0.94\textwidth]{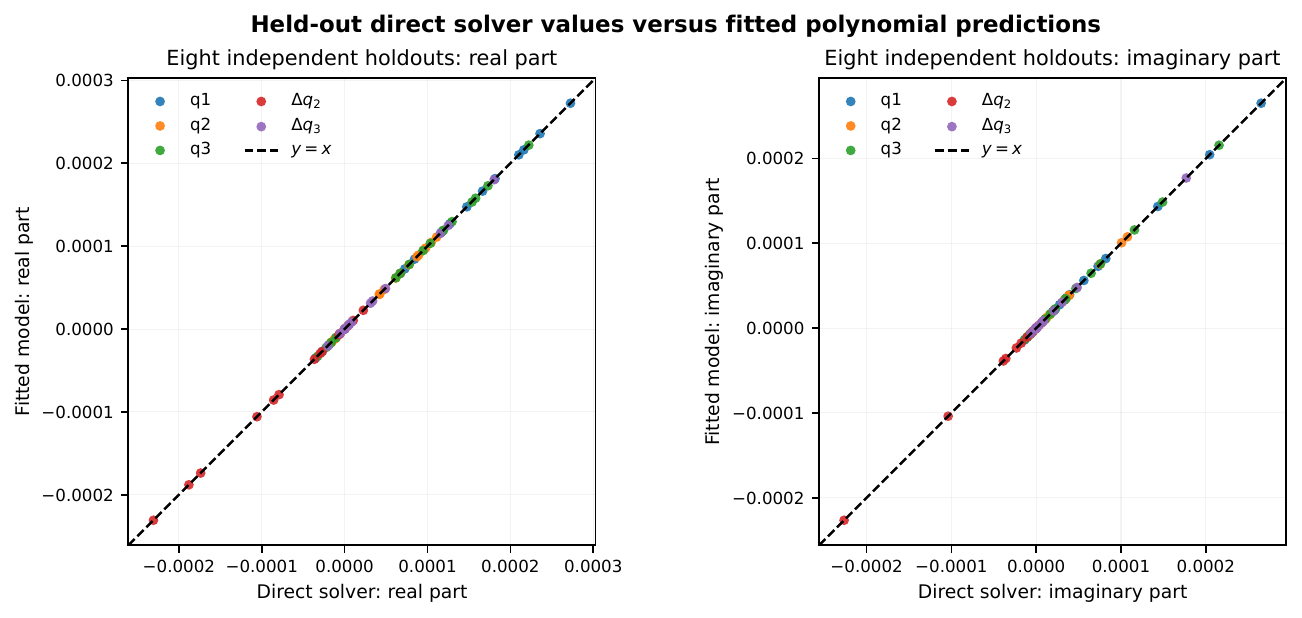}
\caption{Independent validation on eight held-out inputs. Fitted-polynomial predictions are compared with direct finite-cluster solver values for the real and imaginary parts of $q_1$, $q_2$, $q_3$, and the connected increments $\Delta q_2=q_2-q_1$ and $\Delta q_3=q_3-q_2$.}
\label{fig:validation}
\end{figure*}

The numerical calculation retains all five internal levels and directly solves the isolated one-, two-, and three-atom steady states. The finite-cluster treatment is motivated by earlier studies of interacting Rydberg gases and their nonlinear optical response \cite{Sevincli2011CPT,Ates2011EIT,Schempp2010CPT,Stanojevic2013Dispersive,Bai2016FifthOrder,Mu2021GiantKerr}. The single-atom problem is a trace-constrained $25$-component linear system. The pair-connected equation is solved in the $625$-dimensional operator space, and the triplet-connected equation is solved in the $15625$-dimensional space using sparse GMRES with ILU preconditioning and residual refinement. Direct triplet steady-state solutions are sampled as an independent check. The spatial integrals are evaluated using a logarithmic radial grid with Simpson weights and embedded Gauss--Kronrod angular quadrature over the range $r_c=1~\mu$m to $r_{\max}=300~\mu$m. These settings define the finite-range numerical model used for the illustrative propagation results.

All atoms are driven by the same two-component probe. We introduce the dimensionless variables
\begin{equation}
x=\frac{\Omega_{p1}}{\Omega_{\mathrm{ref}}},\qquad
y=\frac{\Omega_{p2}}{\Omega_{\mathrm{ref}}},\qquad
\Omega_{\mathrm{ref}}=0.3~\mathrm{rad}\,\mu\mathrm{s}^{-1},
\end{equation}
where $\Omega_{p1}$ and $\Omega_{p2}$ are the complex probe Rabi frequencies in angular-frequency units. Let $\epsilon=(|x|^2+|y|^2)^{1/2}$. Local charge covariance allows $2$, $6$, and $12$ monomials at total degrees $1$, $3$, and $5$, respectively. The response vector $\boldsymbol q=(\rho_{31},\rho_{32})^T$ is expanded as
\begin{equation}
\boldsymbol q(x,y)=A_1F_1(x,y)+A_3F_3(x,y)+A_5F_5(x,y)+O(\epsilon^7),
\end{equation}
where the ordered monomial vectors $F_1$, $F_3$, and $F_5$ are listed in the Supplementary Material. The coefficient matrices $A_1$, $A_3$, and $A_5$ have dimensions $2\times2$, $2\times6$, and $2\times12$, respectively. We fit $q_1$, $q_2$, and $q_3$ independently and obtain the optical increments from coefficient differences, $\Delta q_2=q_2-q_1$ and $\Delta q_3=q_3-q_2$.

The 32-input design uses four dimensionless probe-amplitude multipliers: $S=0.4,0.7,1.0,1.3$. The reference component amplitude is $\Omega_{p0}=2\pi\times0.0477$ MHz $=0.2997~\mathrm{rad}\,\mu\mathrm{s}^{-1}$, close to $\Omega_{\mathrm{ref}}$. At each setting the total probe intensity is fixed at $|\Omega_{p1}|^2+|\Omega_{p2}|^2=2(\Omega_{p0}S)^2$. Six varied polarization inputs at each setting form the 24-point training set. Two additional equal-component inputs at each setting, with $\Omega_{p1}=\Omega_{p0}S$ and $\Omega_{p2}=\Omega_{p0}S e^{\mathrm{i}\phi}$ for relative phases $\phi=0$ and $\pi/2$, serve as eight independent validation inputs.

Figure~\ref{fig:scaling} shows the directly calculated validation responses over $S\in[0.4,1.3]$. For the equal-component mode with relative phase $\phi=0$, the fitted exponents are $0.9989$ for $q_1$, $2.9973$ for $\Delta q_2$, and $4.9945$ for $\Delta q_3$. The corresponding coefficients of determination exceed $0.999996$.

\begin{table}[t]
\centering
\caption{Relative fitting errors for the 24 training inputs and eight independent validation inputs.}
\label{tab:fit}
\begin{tabular}{lcc}
\toprule
Response & Training error & Validation error\\
\midrule
$q_1$ & $3.26\times10^{-10}$ & $2.08\times10^{-8}$\\
$q_2$ & $4.98\times10^{-7}$ & $2.21\times10^{-5}$\\
$q_3$ & $4.06\times10^{-5}$ & $1.58\times10^{-3}$\\
$\Delta q_2$ & $3.86\times10^{-7}$ & $1.60\times10^{-5}$\\
$\Delta q_3$ & $7.52\times10^{-5}$ & $2.72\times10^{-3}$\\
\bottomrule
\end{tabular}
\end{table}

Figure~\ref{fig:validation} compares the real and imaginary parts of the fitted complex response with the direct finite-cluster solver. On the independent validation set, the largest relative error is $2.72\times10^{-3}$ for $\Delta q_3$ (0.272\%); the corresponding error for $q_3$ is $1.58\times10^{-3}$ (0.158\%).

\begin{figure*}[t]
\centering\includegraphics[width=0.94\textwidth]{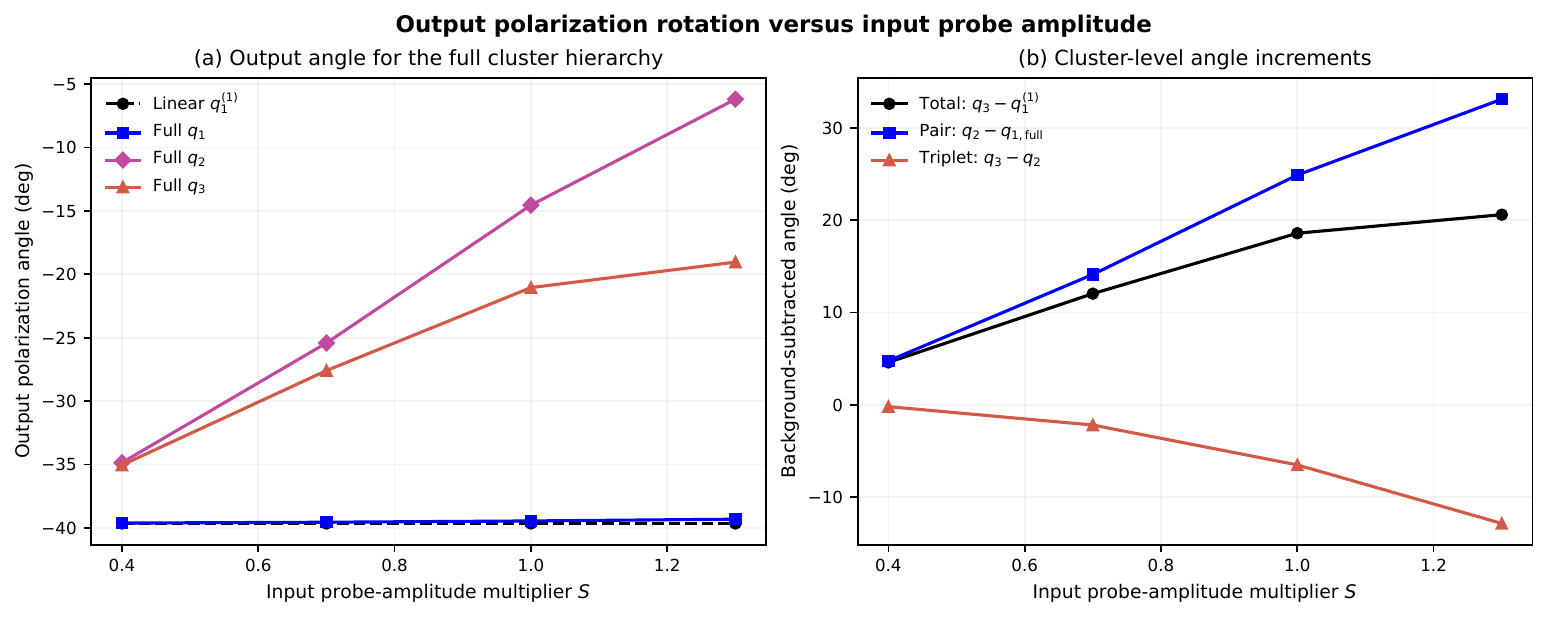}
\caption{Output polarization angle versus input probe-amplitude multiplier. Panel (a) compares the linear single-atom background with the full $q_1$, $q_2$, and $q_3$ responses. Panel (b) shows the total shift from the linear response, the pair increment $\psi(q_2)-\psi(q_{1,\mathrm{full}})$, and the triplet increment $\psi(q_3)-\psi(q_2)$. The pair increment is referenced to the full single-atom polynomial, not the linear background.}
\label{fig:rotation}
\end{figure*}

To connect the local response hierarchy to an observable polarization signal, we propagate the two probe envelopes using the continuous-wave, slowly varying-envelope equations
\begin{equation}
\begin{aligned}
\frac{\partial\Omega_{p,\alpha}}{\partial z}&=\ii\kappa_\alpha q_{\alpha}^{[m]}(\Omega_{p,1},\Omega_{p,2}),\qquad \alpha=1,2,\quad m=1,2,3,\\
\kappa_\alpha&=\frac{\nu|d_{\alpha3}|^2\omega_p}{2\epsilon_0c\hbar}.
\end{aligned}
\label{eq:maxwell}
\end{equation}
Here $\alpha=1,2$ labels the two optical channels, while $m=1,2,3$ labels the single-, pair-, and triplet-cluster truncations; $q_{\alpha}^{[m]}$ is the corresponding channel component of the fitted response vector. The Stokes parameters are $S_0=|\Omega_{p1}|^2+|\Omega_{p2}|^2$, $S_1=2\mathrm{Re}(\Omega_{p1}^*\Omega_{p2})$, $S_2=2\mathrm{Im}(\Omega_{p1}^*\Omega_{p2})$, and $S_3=|\Omega_{p1}|^2-|\Omega_{p2}|^2$, with polarization angle $\psi=\tfrac12\mathrm{atan2}(S_2,S_1)$, following standard polarized-light conventions \cite{BornWolf1999,AzzamBashara1977}.

The propagation uses $\nu=8\times10^{16}$ m$^{-3}$, $\lambda_p=780$ nm, and a medium length of $15$ mm. In the reduced Maxwell propagation, both channels use the same effective dipole moment, $d_{13}=d_{23}=3.584\times10^{-29}$ C m. At the representative input $S=1$, the full $q_3$ propagation trajectory remains within the range of probe-amplitude multipliers used in the fit, $S\in[0.4,1.3]$. This interval describes the amplitude settings used in the fit design; it is not a mathematical domain restriction on the fitted polynomial. Predictions outside the sampled range are polynomial evaluations and have not been independently validated. The calculated output angles are $\psi(q_{1,\mathrm{linear}})=-39.65^\circ$ for the linear single-atom response, $\psi(q_{1,\mathrm{full}})=-39.45^\circ$ for the full single-atom polynomial, $\psi(q_2)=-14.55^\circ$ for the pair-truncated response, and $\psi(q_3)=-21.05^\circ$ for the triplet-truncated response. The single-atom nonlinear correction is $+0.20^\circ$; the pair increment $\psi(q_2)-\psi(q_{1,\mathrm{full}})$ is $+24.91^\circ$, and the triplet increment $\psi(q_3)-\psi(q_2)$ is $-6.50^\circ$. The total shift relative to the linear background is $+18.60^\circ$. Equivalently, the pair and triplet increments sum to $+18.40^\circ$ relative to the full single-atom response; the remaining difference is due to the single-atom nonlinear correction.

The output polarization angle as a function of the input probe-amplitude multiplier is shown in Fig.~\ref{fig:rotation}.

To isolate the role of the symmetry-allowed target orders, we also propagate diagnostic responses in which the pair increment is restricted to its cubic coefficient block and the triplet increment to its quintic coefficient block. These order-isolated curves are diagnostics, not replacements for the full $q_1$, $q_2$, and $q_3$ responses. At $S=1$, the full pair increment and its third-order-only diagnostic are $24.91^\circ$ and $25.00^\circ$, respectively; the full triplet increment and its fifth-order-only diagnostic are $-6.50^\circ$ and $-6.05^\circ$.

Local symmetry constrains not only the form of nonlinear optical response but also the connected many-body sectors that can contribute. For the specified driven-dissipative Rydberg--EIT dynamics, an $n$-atom connected root-optical response is forbidden below probe-amplitude order $2n-1$. Hence coefficients through fifth order receive contributions only from single-, pair-, and triplet-connected clusters, without imposing weak interatomic interactions. Our finite-cluster calculations support this hierarchy, and illustrative propagation reveals partial cancellation of pair-induced polarization rotation by the triplet sector. The selection rule applies to optical Taylor coefficients; it does not establish convergence of a finite-density cluster expansion at arbitrary probe strength. Its combination with density and intensity dependence suggests a route toward identifying optically distinguishable higher-order connected contributions.

\begin{acknowledgments}
This work is supported by the Key Special Project of the National Key Research and Development Program of China (Grant No. 2025YFF0515201), the Joint Fund for Quantum Major Research Plan of the National Natural Science Foundation of China (Grant No. U25D8014), and the Natural Science Foundation of Chongqing (Grant No. CSTB2026NSCQ-GPX1195).
\end{acknowledgments}

\bibliographystyle{apsrev4-2}
\bibliography{references}
\end{document}